# Analysis of liver cancer detection based on image processing

**Mahmoudreza Moghimhanjani\* , Ali Taghavirashidizadeh**

1. Department of Biomedical Engineering, Islamic Azad University,Central Tehran Branch(IAUCTB), moghim.mahmoodreza@yahoo.com
2. Department of Electrical and Electronics Engineering, Islamic Azad University,Central Tehran Branch(IAUCTB), ali.taghavi.eng@iauctb.ac.ir

**Abstract**
Medical imaging is the most important tool for detecting complications in the inner body of medicine. Nowadays, with the development of image processing technology as well as changing the size of photos to higher resolution images in the field of digital medical imaging, there is an efficient and accurate system for segmenting this. Real-world images that for a variety of reasons have poor heterogeneity, noise and contrast are essential. Digital image segmentation in medicine is used for diagnostic and therapeutic analysis, which is very helpful for physicians. In this study, we aim at liver cancer photographs, which aim to more accurately detect the lesion or tumor of the liver because accurate and timely detection of the tumor is very important in the survival and life of the patient.The aim of this paper is to simplify the obnoxious study problems related to the study of MR images. The liver is the second organ most generic involved by metastatic disease being liver cancer one of the prominent causes of death worldwide. Without healthy liver a person cannot survive. It is life threatening disease which is very challenging perceptible for both medical and engineering technologists. Medical image processing is used as a non-invasive method to detect tumours. The chances of survival having liver Tumor highly depends on early detection of Tumor and then classification as cancerous and noncancerous tumours. Image processing techniques for automatic detection of brain are includes pre-processing and enhancement, image segmentation, classification and volume calculation, Poly techniques have been developed for the detection of liver Tumor and different liver toM oR detection algorithms and methodologies utilized for Tumor diagnosis. Novel methodology for the detection and diagnosis of liver Tumor.

## 1. Introduction
Liver cancer is a silent and dangerous disease that has no symptoms at an early stage. The first symptoms of liver cancer usually appear as the disease progresses, but there are vague and general symptoms such as fatigue, anorexia, nausea, and so on that most of the time the person does not pay attention to and when he or she is diagnosed with a disease that there is little work left to do. The following is what liver cancer is all about and how to prevent and treat liver cancer. What is Liver Cancer? It is a type of cancer that begins in the liver. Some cancers occur outside the liver and then spread to the area. However, only cancers that start from the liver itself are known as liver cancers.

The liver, located at the bottom of the right lung below the chest, is one of the largest organs in the human body. The liver has many functions such as removing toxins from the body and is essential for life. Liver cancer is the presence of malignant liver tumors inside or on the liver. Every year in the United States, about 3,000 men and 6,000 women are diagnosed with liver cancer, of which about 2,000 are men and 6,000 are women. Liver cancer is one of the most common malignant diseases worldwide. One of the most common and best imaging modalities to diagnose liver diseases such as hepatocellular carcinoma is CT [1-5]. CT data has a high level of noise, and because the contrast between the tumor and the main liver is low, it is difficult to detect and reliably map the tumor. In addition to the limitations of the imaging procedure, liver tumor segmentation will be complicated because of the variety of tumor size and structure and because tumors can be present virtually anywhere in the liver. This thesis attempts to identify the tumor segment in images of patients with liver tumors. The proposed procedure for tumor segmentation is performed in three stages. Pre-processing will take place on the images first. Then, using a histogram of images and morphological operations and using anatomical information about the shape and location of the liver, an initial search area is obtained that will eventually search for the exact border of the liver. In the third stage, the obtained area is divided into blocks and the tumor segment will be obtained using SOM neural network[6-8]. The results of this algorithm show accuracy of about 2%.

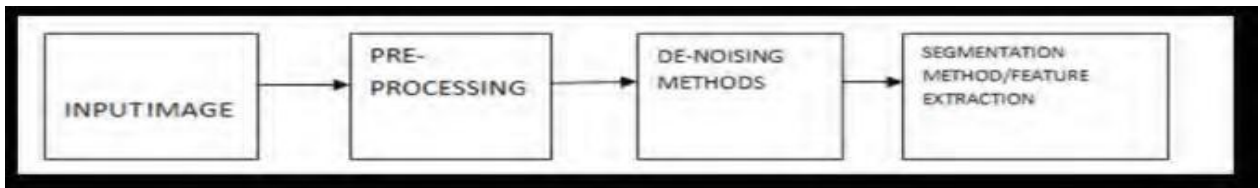

**Fig. 1. Basic Block Diagram**

**2. Liver cancer diagnosis:**
Tests and methods used to diagnose liver cancer include:

Blood tests:

Blood tests may show abnormalities in liver function.

Imaging Tests:

Your doctor may recommend imaging tests, such as ultrasound, computed tomography, and MRI.

A sample of liver tissue for examination:

Your doctor may recommend that you remove a piece of liver tissue for laboratory testing to make a definitive diagnosis of liver cancer.
Determine the extent of liver cancer:

After diagnosing liver cancer, your doctor will do things to determine the extent (stage) of the cancer. Visual tests help determine the size and location of the cancer and whether it has spread. Imaging tests are important here.

There are different ways to diagnose liver cancer. One method uses numbers I to four, and the other uses letters A to D.

Stage 4 and stage D show the highest rate of liver cancer with the worst predictor.

### 3. PROPOSED METHODOLOGY AND ANALYSIS
A. Image Enhancement
Image enhancement is image pre-processing stage. The purpose of the process of image enhancement is to improve
The image quality for the human eye. This process is also require to provide a better input image for further processing, so that the result of the image after processing all the stages contains less errors.
The image enhancement technique is divided into two parts which are spatial domain technique and frequency domain technique. In spatial domain technique the value of the pixel is changed with respect to the requirement whereas the frequency domain technique deals with the rate of change of pixels which are changing due to spatial domain. It cannot be determined that what type of technique is good for image enhancement. There are many techniques for image enhancement technique out of which we have use ostu's method.
1) Otsu's method
This method uses clustering based technique shows in Fig. 2 in other words, it convert greyscale image into binary
image. It assumes that the image contain two level of pixel which are foreground pixel and background pixel (bi-modal histogram). It calculate the optimum threshold by separating two classes. The result gives minimum combined-spread and maximum inter-class variance. The ostu's method Fig. 3 can roughly said to be one-dimensional method. The ostu's method search for the threshold which minimizes the interclass variance.

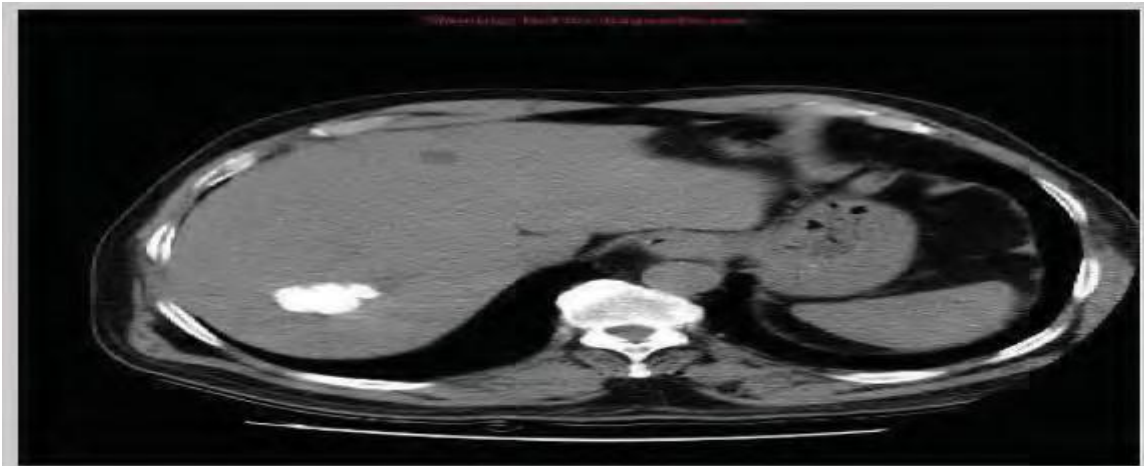

**Fig. 2. Original image**

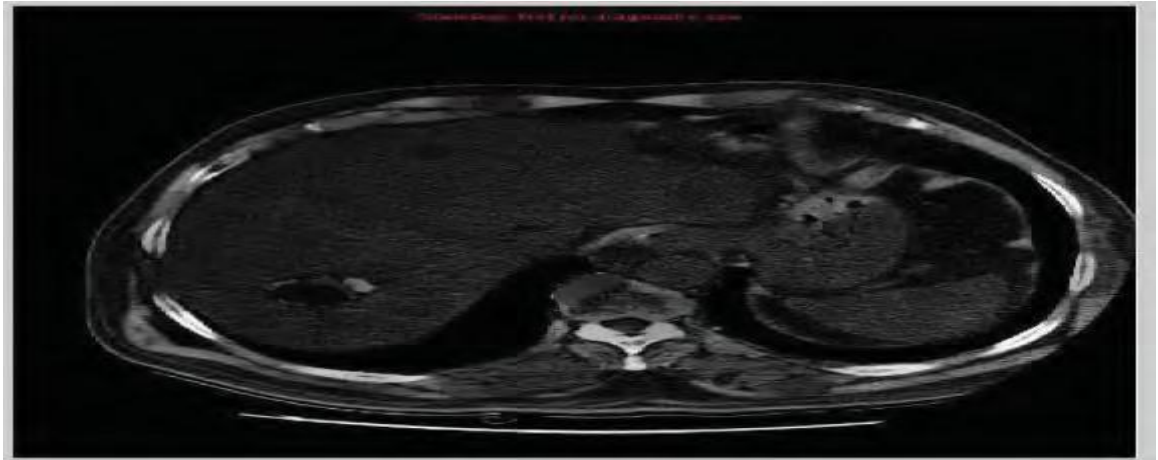

**Fig. 3. Otsu's Transformation**

B. Image Segmentation

It is an important process for many task in image processing. Many of the important techniques like image description and image recognition are depend on the image segmentation. The process of segmentation divides an image into region or object. The image processing segments 2D image and it has numerous application in the field of medical. This may include visualization, estimation of volume of the interest object, detecting abnormalities like tumors, polyps etc. and tissue qualification and much more. The objective of the process of segmentation is to make the image more useful by changing the representation and simplifying the image due to which it will be easier to analyze the image. This process is use for detecting the boundaries and objects of an image. More precisely we could defined image segmentation as the process of assigning a name or label to the each pixel of that particular image which share certain visual characteristics. The result of a segmentation of an image is basically the entire image which will be formed by combining each segmented parts. The characteristics and the properties will also be same like contour, intensity or texture. However, the adjacent segments contains different characteristics. The segmentation process has two basic properties on intensity values namely discontinuity and similarity. The discontinuity property partitioned the image into different regions for example edges of an image. The similarity property partitioned the image into regions which has similar predefined criterion.The gradient magnitude shows in Fig. 4.

1) Marker-Controlled Watershed Segmentation Approach

Marker-Controlled Watershed Segmentation process enhance the region which indicate the presence of the required object. The location which are extracted by this process are then set to the minimum position within the same topological surface. The watershed algorithm is applied afterwards. Separating objects of an image is one of the difficult method which watershed segmentation makes it easier. Watershed Segmentation Approach is of two types: External associated with Background and Internal associated with the object of interest.The watershed transformation of the Gradient image shows in Fig. 5.

Image segmentation use watershed transform to locate the foreground and background object location. Opening and Closing Reconstruction results shows in Fig. 6 and Fig. 7. This locate the "catchment basins" and "watershed ringe line" of an image by treating its surface with light pixel classified as high and low pixel classified as low. The Fig. 8 and Fig. 9 shows the results of watershed segmentation results:

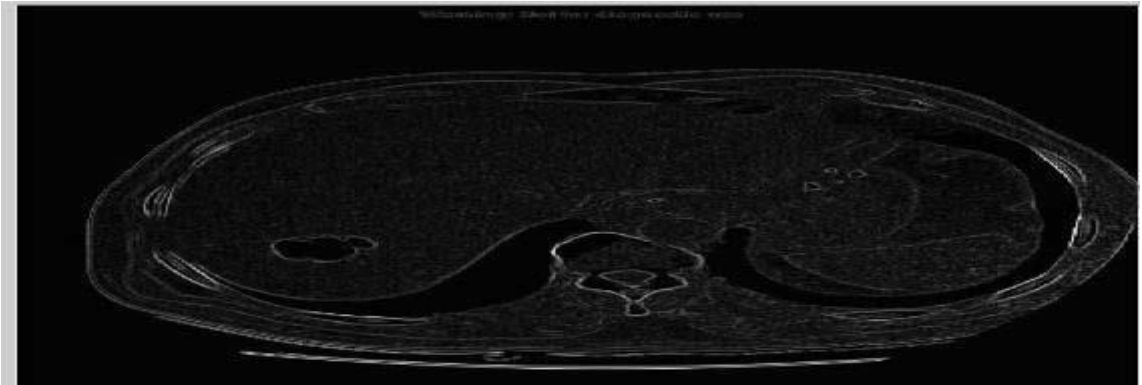

**Fig. 4. Gradient magnitude**

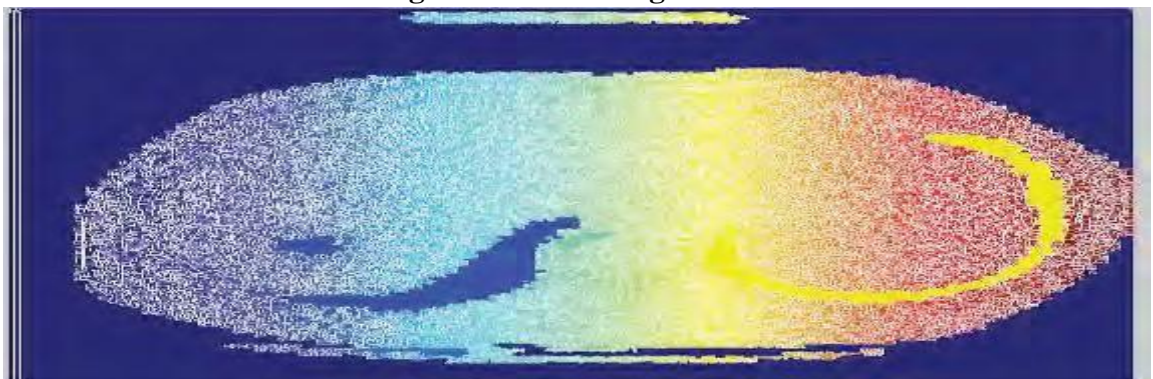

**Fig. 5. Watershed Transform of the gradient magnitude image**

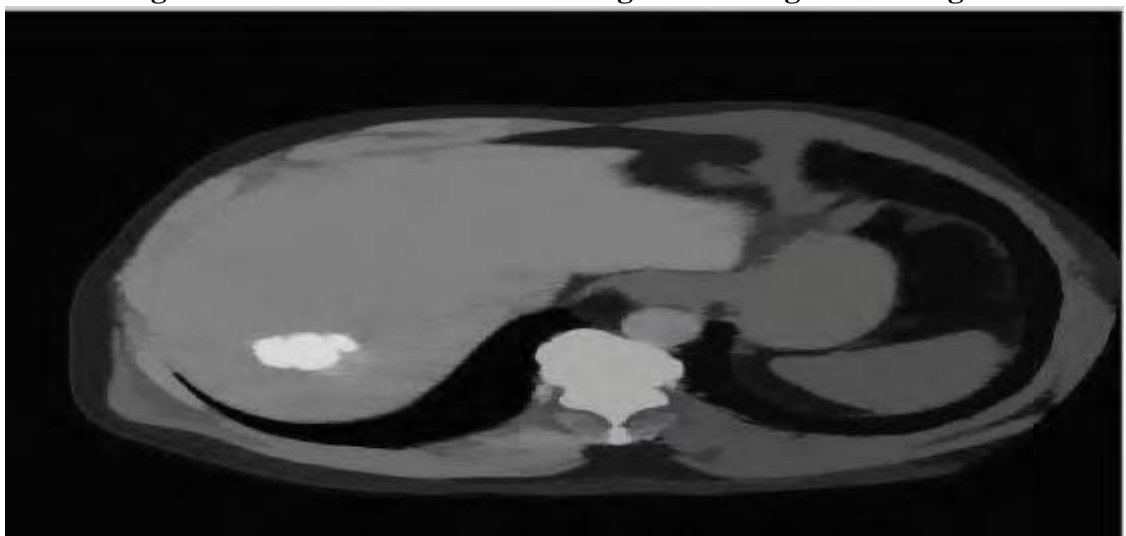

**Fig. 6. Closing by reconstruction**

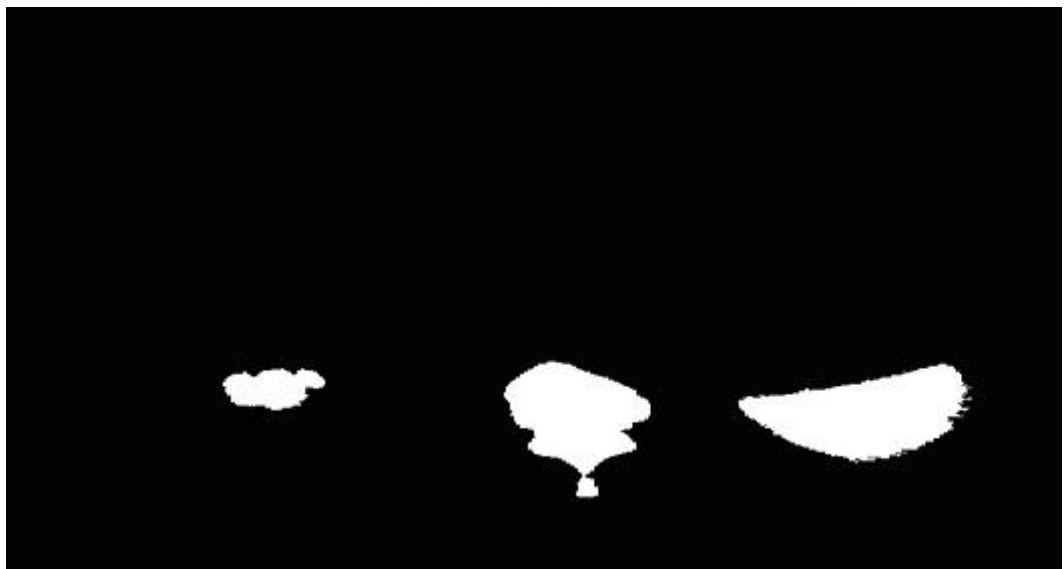

**Fig. 7. Regional maxima of opening closing by reconstruction**

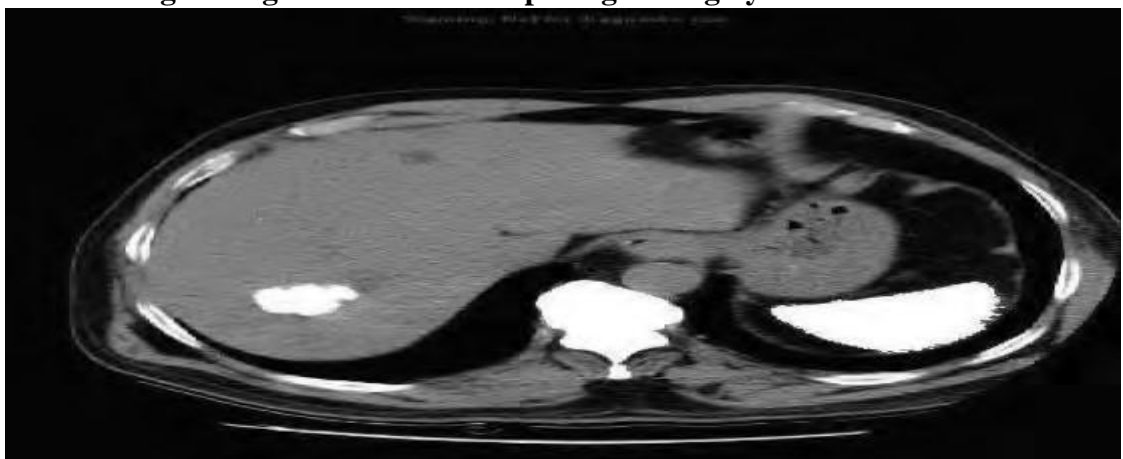

**Fig . 8. Regional maxima superimposed on original image**

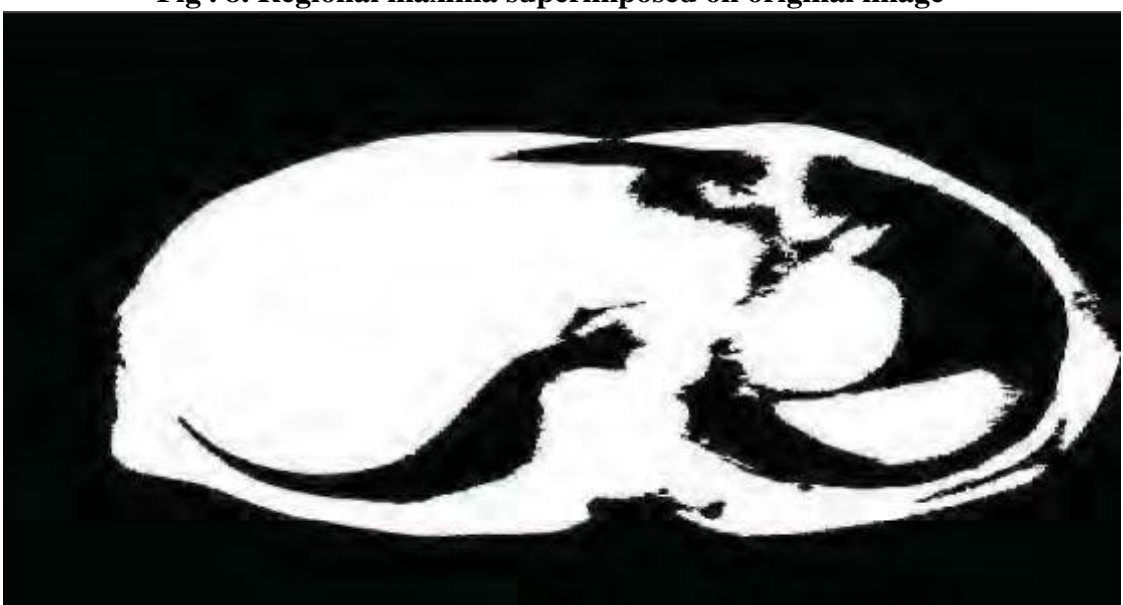

**Fig 9. Threshold opening-closing by reconstruction**

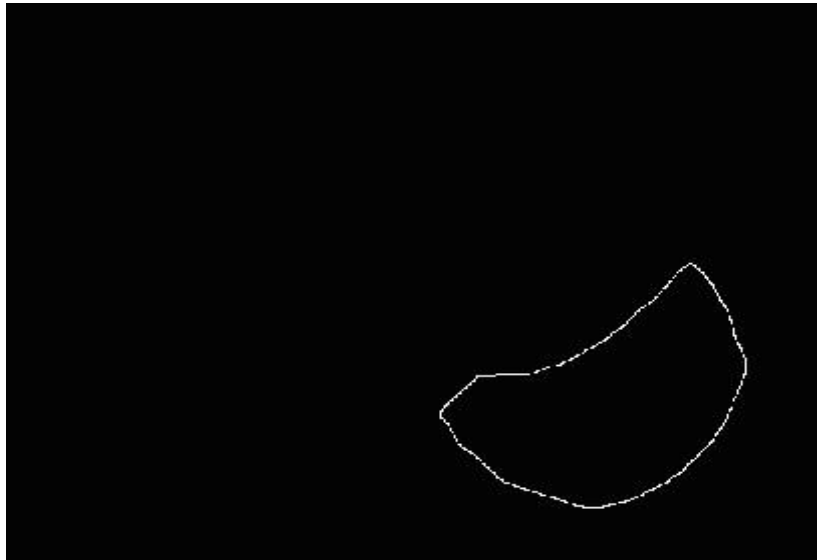
**Fig 10. Watershed ridge line**

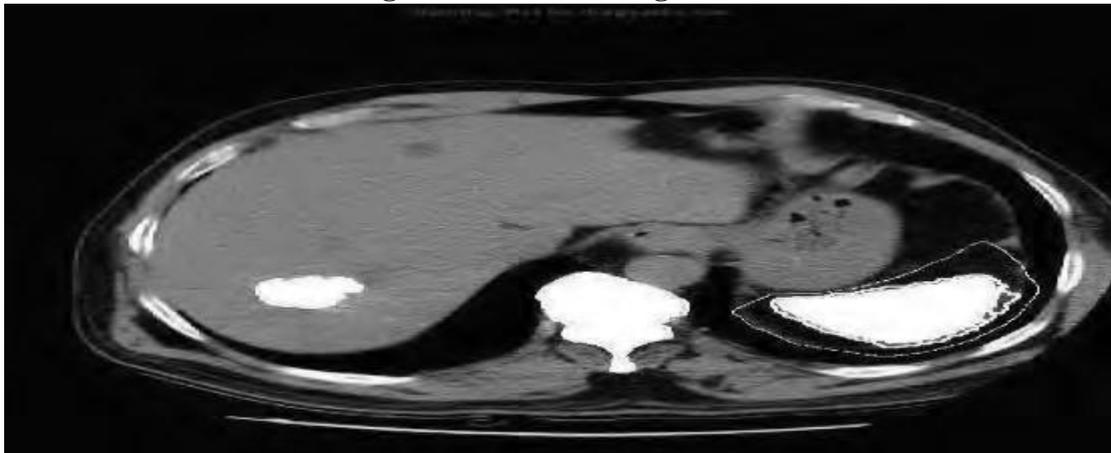
**Fig. 11. Markers and object boundaries superimposed on original image**

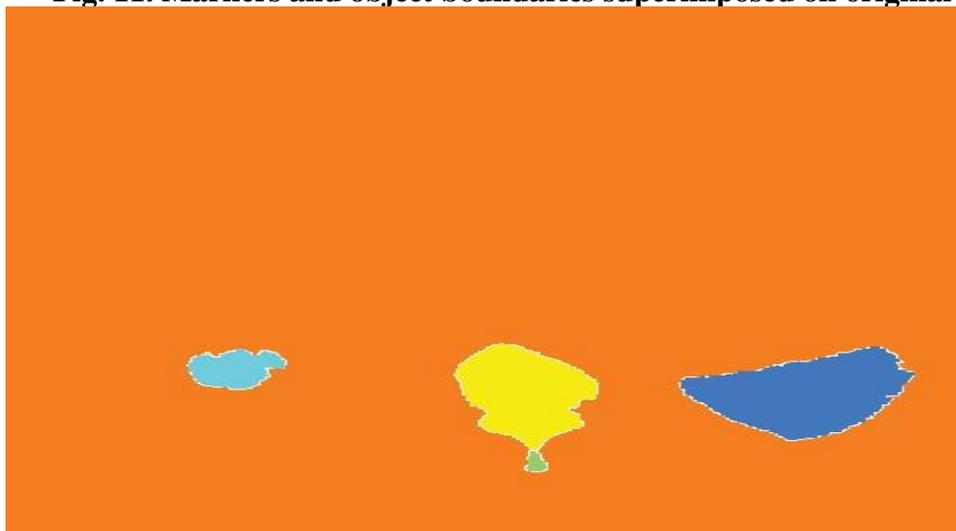
**Fig. 12. Watershed label matrix**

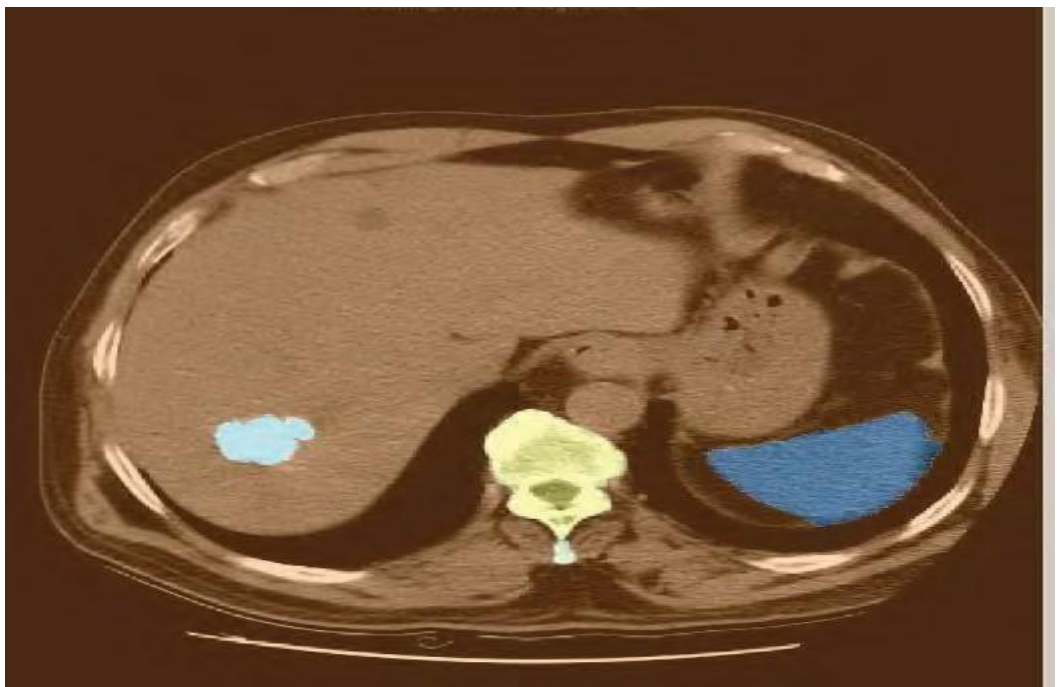

**Fig. 13. Superimposed transparently on original image**

C. Feature extraction

Image feature extraction is one of the most important technique of image processing Watershed ridge and boundaries images is shown in Fig. 10 and Fig. 11. It uses different techniques and algorithm to isolate and detect various shapes and portions of the image. The watershed label matrix and Superimposed image is shown in Fig. 12 and Fig. 13 There are numerous techniques to apply this to the image.Wavelet transform is one of the tool for feature extraction. The wavelet transform has a characteristic of analyzing the image with varying unit of resolution and has multi resolution analytic property [9]. The wavelet transform is better than Fourier transform and short time Fourier transform as it preserves both time and frequency as in Fourier transform it discard the time.

**4. CONCLUSION**

Different MRI Images where acquired from the internet, basic Ostu preprocessing technique was used, for segmentation Marker-Controlled Watershed Segmentation was used and it was observed that for a few images' segmentation was done correctly, so our future works includes creating a GUI and enabling a single click feature extraction using wavelet transform, with the accuracy.

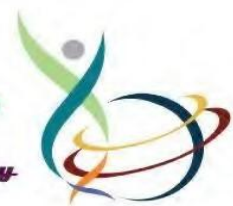

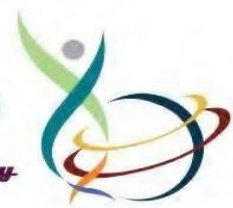